\documentclass[12pt]{article}
\renewcommand{\baselinestretch}{1.2}
\usepackage[utf8]{inputenc}

\usepackage{indentfirst}
\usepackage{amsmath}
\usepackage{amssymb}
\usepackage{amsfonts}
\usepackage{amscd}
\usepackage{amsbsy}
\usepackage{amsthm}
\usepackage{latexsym}
\usepackage{graphicx,color} 
\usepackage[dvipsnames]{xcolor}
\usepackage[colorlinks]{hyperref}
\hypersetup{linkcolor=blue,citecolor=blue,urlcolor=blue}

\def\nn{\nonumber}       
\def\beq{\begin{eqnarray}}
\def\eeq{\end{eqnarray}}
\def\ln{\,\mbox{ln}\,}

\def\tr{\,\mbox{tr}\,}

\def\diag{\,\mbox{diag}\,}
\def\Tr{\,\mbox{Tr}\,}

\def\al{\alpha}
\def\be{\beta}

\def\ga{\gamma}
\def\de{\delta}
\def\vp{\varepsilon}
\def\ep{\epsilon}

\def\ka{\kappa}
\def\la{\lambda}
\def\na{\nabla}
\def\pa{\partial}

\def\rh{\rho}
\def\si{\sigma}

\def\ph{\varphi}

\def\th{\theta}

\def\Ga{\Gamma}
\def\De{\Delta}

\usepackage[symbol]{footmisc} 
\usepackage{float} 
\usepackage{cite}  
\usepackage{titlesec}

\usepackage{ulem} 

\titleformat*{\section}{\large\bfseries}
\titleformat*{\subsection}{\normalsize\bfseries}

\begin{document}

\begin{center}
\renewcommand*{\thefootnote}{\fnsymbol{footnote}} 
{\Large \bf
On the new way of symmetry breaking in scalar
\\
QED and the one-loop renormalization}
\vskip 6mm

{\bf Lucas Ducastelo de C. C. Lima}
\hspace{-1mm}\footnote{E-mail address: \ lucasducastelo@yahoo.com.br}
\quad
and
\quad
{\bf Ilya L. Shapiro}
\hspace{-1mm}\footnote{E-mail address: \ ilyashapiro2003@ufjf.br}
\vskip 6mm

Departamento de F\'{\i}sica, ICE, Universidade Federal de Juiz de Fora,
\\
Juiz de Fora, 36036-900, Minas Gerais, Brazil
\end{center}
\vskip 2mm
\vskip 2mm


\begin{abstract}

\noindent
It is well known that a single real scalar field does not allow gauge
coupling to the Abelian vector field. Using the complex scalar model
as a starting point, we construct the Abelian gauge model with two
real scalars. The gauge transformations for the scalars look
different (albeit equivalent) from the conventional sQED. One can
show that the invariant action cannot be extended by adding more
scalars. On the other hand, in the theory with two real scalars,
spitting their masses, or the scalar self-coupling constants, or the
nonminimal parameters of scalar-curvature interaction, we arrive
at a qualitatively new way of gauge symmetry breaking. Using the
Schwinger-DeWitt technique, we explore the one-loop renormalization
of this new model in curved spacetime.
\vskip 3mm

\vskip 3mm


\end{abstract}
\setcounter{footnote}{0} 
\renewcommand*{\thefootnote}{\arabic{footnote}} 

\section{Introduction}
\label{sec1}

Gauge symmetry cannot be defined for a theory with single real
scalar field, as it is impossible to couple gauge field $A_\mu$ to
such a scalar. On the other hand, the theory of a complex scalar
field plays a fundamental role in describing gauge interactions of
spin-zero particles, especially the Higgs boson. Following one of the
exercises in the recent textbook \cite{OUP}, we note that the Abelian
gauge model of complex scalar can be reformulated in terms of two
real scalar fields. In contrast, the rule of the gauge transformation
gets modified to preserve gauge invariance. The scheme is not
affected by the presence of an external gravitational field. This is
a potentially interesting detail because the models with two real
scalars have applications in the inflationary cosmological models
(see, e.g.,
\cite{Liddle1998,Wands,PetersonTegmark2011} and
more recent consideration for the combined Higgs-$R^2$ inflation
in \cite{Heatal}).
Typically, these
models of ``assisted inflation'' are characterized by a strong mass
hierarchy. Furthermore, we note that the generalization of the new
representation to the non-abelian case looks straightforward,
regardless it lies beyond the scope of the present work. In
particular, one can construct such representation for the $SU(2)$
doublet describing the Higgs field.

Looking at the situation from the perspective of the Higgs field, one
can ask whether the representation of the well-known gauge theory
in terms of the two real scalar fields can lead to an alternative
mechanism for gauge symmetry breaking. To investigate this
possibility, we consider an action in which the two real scalar
fields acquire different masses or/and different nonminimal couplings
to the scalar curvature, or get a symmetry violation from the
self-interaction of the two scalars. All these possibilities lead to the
\textit{explicit} symmetry breaking, in contrast to the known ways
(spontaneous or dynamical) of symmetry breaking, which can be
used to generate a non-zero mass for the gauge vector field.

At very high energies, typical for inflation, the quantum effects may
be relevant. It is worth mentioning the recent investigation of such
effects in the multiscalar model \cite{vZ2}, including curved-space
effective potential, derivative expansion in the one-loop effective
action, and the analysis of low-energy decoupling. It would be
interesting to check whether the explicit symmetry breaking described
above might lead to a consistent quantum theory. In particular, one
can ask whether the theory with the new type of
symmetry breaking may be renormalizable. To address this question,
we evaluate the one-loop renormalization for the theory with broken
symmetry in curved spacetime with a metric $g_{\mu\nu}$.
The renormalizability beyond one-loop level is an unsolved issue,
which is not easy to analyse because there is no symmetry which
would prevent the non-symmetric counterterms of the new type to
emerge in higher loops. We present this part of our considerations
in Appendix B.

The manuscript is organized as follows. Sec.~\ref{sec2} describes
the classical gauge model with two scalar fields and introduces the
explicit symmetry breaking.  Sec.~\ref{sec3} presents the discussion
of the possible extension of the Abelian model to the case of
extra scalar fields. In Sec.~\ref{sec4} we derive the
one-loop counterterms and show that the theory with broken symmetry
is renormalizable, at least at this level. Sec.~\ref{sec5} describes
the renormalization group equations and explores the IR limit of the
effective charges corresponding to the broken symmetry. Finally,
in the last Sec.~\ref{sec6}, we draw our conclusions and outline
possible extensions of the present article.

\section{Classical action and gauge invariance}
\label{sec2}

Our starting point is the Abelian gauge theory for a complex scalar
field,
\beq
S_c \,\,=\,\,
\int_x \Big\{g^{\mu\nu}(D_\mu\phi)^* \,(D_\nu\phi)
- m^2 \phi\phi^*
+ \xi R \phi\phi^*
- \frac{\la}{12}\,(\phi\phi^*)^2 \Big\},
\label{acaocomplex}
\eeq
where $m$ is the mass of the scalar field, $\xi$ is the parameter of
nonminimal scalar-curvature interaction, $\la$ is the self-interaction
coupling constant, the covariant derivatives are
\beq
&&
D_\mu\phi = \na_\mu \phi - igA_\mu\phi
\qquad
\mbox{and}
\qquad
(D_\nu\phi)^* = \na_\nu \phi^* + igA_\nu\phi^*
\eeq
and, finally, we use the abbreviations
\beq
(\na \ph)^2 = g^{\mu\nu} \pa_\mu \ph \pa_\nu \ph ,
\quad
\int_{x,n} =  \int d^nx \sqrt{-g}
\quad
\mbox{and}
\quad
\int_x  =  \int_{x,4}.
\label{int}
\eeq
The field transformations ensuring
the gauge invariance of Eq.~\eqref{acaocomplex} are
\beq
{A'}_\mu \;=\; A_\mu + \pa_\mu f
\quad
\mbox{and}
\quad
\phi' \;=\; e^{igf}\phi ,
\quad
\mbox{with}
\quad
f = f(x).
\eeq
Let us express the complex field in terms of two real scalars
$\ph$ and $\chi$ as
\beq
\phi \;=\; \frac{1}{\sqrt{2}} (\ph \, + \, i \chi).
\eeq
Replacing this decomposition into \eqref{acaocomplex},
we arrive at
\beq
&&
S_r \,\,=\,\,
 \frac{1}{2}\,\int_x\,\Big\{
    (\na \ph)^2 + (\na \chi)^2
    + g^2 A^2 \big(\ph^2 + \chi^2\big)
    - 2 g A^\mu  \big(\ph\na_\mu \chi - \chi \na_\mu \ph\big)
\nn
\\
&&
\qquad\qquad
    -\,\,
    m^2\big( \ph^2 +\chi^2\big)
    + \xi R \, \big(\ph^2 + \chi^2 \big)
    -\frac{\la}{12}\,\big(\ph^2+\chi^2\big)^2 \Big\} \, .
\label{acao2esc}
\eeq
The transformations of scalars providing the gauge invariance
in Eq.~\eqref{acao2esc} have the form
\beq
&&
\ph' \;=\; \ph \cos(gf) - \chi \sin(gf)\, ,
\nn
\\
&&
\chi' \;=\; \ph \sin(gf) + \chi \cos(gf)\,.
\label{rotaphichi}
\eeq
When this transformation is replaced into the action (\ref{acao2esc}),
all $\sin(gf)$ and $\cos(gf)$ cancel out and we meet the expected
invariance.

One can break the gauge symmetry by introducing a mass-splitting
in the scalar sector, or/and by choosing different nonminimal
couplings to the scalar curvature, or introduce the symmetry
violation in the scalar self-interaction sector,
\begin{align}
    S_{bs}
    \, &= \int_x\,\Big\{
    \frac{1}{2} (\na \ph)^2 + \frac{1}{2} (\na \chi)^2
    + \frac{1}{2} g^2A^2 (\ph^2 + \chi^2)
    - g A^\mu (\ph \na_\mu \chi - \chi \na_\mu \ph)
    \nonumber
\\
&
\quad
- \, \frac{1}{2} m^2 \ph^2 - \frac{1}{2} M^2 \chi^2
    + \frac{1}{2} \xi R \ph^2 + \frac{1}{2} \Xi R \chi^2
    - \frac{\la}{24} (\ph^4+\chi^4) - \frac{\la_{12}}{12} \ph^2\chi^2
\Big\}.
\label{acaobs}
\end{align}
It is easy to see that $m$ and $\xi$ are associated with the
scalar field $\ph$, while $M$ and $\Xi$ correspond to $\chi$.
Furthermore, $\la$ is the self-interaction coupling constant, and
$\la_{12}$ is a new coupling representing one more way of
the explicit symmetry breaking. In the limit
\beq
M \,\longrightarrow\,m ,
\quad
\Xi \,\longrightarrow\,\xi ,
\quad
\la_{12} \,\longrightarrow\,\la ,
\label{back}
\eeq
we come back to the theory (\ref{acao2esc}), that is equivalent to
the original (\ref{acaocomplex}).

\section{More real scalars?}
\label{sec3}

Before we start the analysis of the quantum theory with explicit
breaking of
gauge symmetry (\ref{acaobs}), let us try to answer a natural
question about the possible extension of the symmetric model
(\ref{acao2esc}). Is it possible to add one more scalar in this
action, maintaining the Abelian gauge symmetry? If true, this would
mean a new realization of this symmetry since the new model would
not be equivalent to the complex scalar theory (\ref{acaocomplex}).
In the rest of this section we try to argue why this possibility does
not work.

Let us introduce another scalar field $\th$ in the action
(\ref{acao2esc}), with the same vector field. The last means, for
each pair of scalars, the charge $g$ is the same. Then the
candidate action can be cast in the form
\begin{align}
S_{\ph\chi\theta} \,\,
&=\,\, S_{\ph\chi} + S_{\chi\th} + S_{\th\ph} + S_{int}
\nn
\\
\,&=\, \frac{1}{2}\,\int_x\,\Big\{
(\na \ph)^2 + (\na \chi)^2 + (\na \theta)^2
+ g^2 A^2\, \big(\ph^2 + \chi^2 + \theta^2\big)
\nn
\\
&
\quad
-\,\,
2g A^\mu \, \big(\ph\na_\mu \chi - \chi \na_\mu \ph
+ \chi\na_\mu \theta - \theta \na_\mu \chi
+ \theta\na_\mu \ph - \ph\na_\mu \theta\big)
\nn
\\
&
\quad
-\,\,
\big(m^2 - \xi R\big) \,\big( \ph^2 + \chi^2 + \theta^2\big)
\,-\,  \frac{\la}{12}\,\big(\ph^2+\chi^2 +\theta^2\big)^2\Big\} \, .
\label{S3scals}
\end{align}
Setting one of the scalars to zero, we arrive at the action
(\ref{acao2esc}), equivalent to the one of complex scalar
(\ref{acaocomplex}). Without this, there is a qualitatively
new theory.

It is certainly possible to have field transformations of the form
(\ref{rotaphichi}) with $\th'=\th$, for any couple of the scalar
fields, such that the respective actions $ S_{\ph\chi}$,
$S_{\chi\th}$, or $S_{\th\ph}$ possess gauge invariance.
The explicit form of the new transformations is
\begin{equation}
\label{LT2}
\begin{aligned}
& \chi' \;=\; \chi \cos(gf) - \th \sin(gf)\, ,
\\
& \th' \;=\; \chi \sin(gf) + \th \cos(gf),
\end{aligned}
\end{equation}
with  $\ph'=\ph$, and
\begin{equation}\label{LT3}
\begin{aligned}
& \th' \;=\; \th \cos(gf) - \ph \sin(gf)\, ,
\\
& \ph' \;=\; \th \sin(gf) + \ph \cos(gf)\,,
\end{aligned}
\end{equation}
with  $\chi'=\chi$, for each of the respective actions.

The question is whether it is possible for the action with
all three scalars to have such an invariance? As far as we can
see, the answer to this question is negative.
As an example, consider (\ref{rotaphichi}). The action
$S_{\ph\chi}$ and the last two terms in (\ref{S3scals}) are
invariant. However, the mixed scalar-vector term has an
element transforming in the odd way, as
\beq
&&
\th' \na_\mu \chi' - \chi' \na_\mu \th'
+ \ph' \na_\mu \th' - \th' \na_\mu \ph'
\,=\,
\big[ \cos(gf) + \sin(gf)\big]
\big( \th \na_\mu \chi - \chi \na_\mu \th  \big)
\nn
\\
&&
\qquad
+\,
\big[\sin(gf) -  \cos(gf)\big]
\big( \th \na_\mu \ph - \ph \na_\mu \th \big)
\,+\,
(\na_\mu f) \big[ \cos(gf) - \sin(gf)\big]\th \chi
\nn
\\
&&
\qquad
+\,\,
(\na_\mu f) \big[ \cos(gf) + \sin(gf)\big]\th \ph \,.
\label{oddtrans}
\eeq
In this case, $\sin(gf)$ and $\cos(gf)$
do not cancel and this cannot be compensated by the
transformation of the vector field.

Similar situation takes place if we introduce more scalar fields.
One can easily provide the symmetry in the free action of each
couple of fields and in the mass, nonminimal and interaction
sectors, but not in the
whole action of $n$ scalar fields. In this sense, the real fields
representation (\ref{acao2esc}) of the gauge invariant action
of charged scalar (\ref{acaocomplex}) does not admit an
extension without introducing non-abelian gauge symmetry.

\section{One-loop divergences}
\label{sec4}

In this section, we compute the one-loop divergences for a theory
with two scalar fields possessing an Abelian gauge symmetry.
\beq
S
\,\, = \,\,S_{bs} \,\,-\,\, \frac{1}{4} \int_x F_{\mu\nu}^2\,.
\label{acaoclas2}
\eeq

Using the background field method, we split
\beq
\ph \to \ph' \,=\, \ph + \rh, \qquad
\chi \to \chi' \,=\, \chi + \si,
\qquad
A_\mu \to A_\mu' \, = \,A_\mu + B_\mu,
\label{campofundo}
\eeq
with $\rh$, $\si$, and $B_\mu$ are quantum fields. The divergences
are defined by the part of the action that is bilinear in the quantum
fields.

The action of the gauge sector can be rewritten as
\begin{equation}
- \frac{1}{4} \int_x F_{\mu\nu}^2
\, = \,
 \frac{1}{2}\int_x B^\mu
 \big(\delta_\mu^\nu \Box - R_\mu^\nu\big) B_\nu
 \,+\, \frac{1}{2}\int_x \big(\na_\mu B^\mu\big)^2
\label{acaoB}
\end{equation}
and has to be extended by adding a gauge fixing term in the
framework of the Faddeev-Popov method.  The simplest option is
\beq
S_{gf}
\,=\, - \,\frac{1}{2} \int_x \big(\nabla_\mu B^\mu\big)^2\,.
\label{Sgf}
\eeq

{At this point, it is worth making an important observation. The
theory with explicitly broken symmetry is not gauge invariant and
hence the use of Faddeev-Popov method is not allowed.
However, at least in the Abelian theory (\ref{acaobs}) things may
be different because the symmetry breaking occurs only in the
scalar sector. In the  Faddeev-Popov approach, the gauge
transformation is performed in the integral (see, e.g., \cite{OUP})
\beq
\De^{-1}\,=\,\int d\vp\,\, \de\big(b - l \big)
\label{FPtrans}
\eeq
over the gauge transformation parameter $\vp$, where $b$ is the
gauge fixing condition and $l$ is an arbitrary function. The integral
is taken by the change of variables
\beq
db \,=\, \Big(
       \frac{db}{dB^\mu}\,\frac{dB^\mu}{d\vp}
\,+\, \frac{db}{d\rho}\,\frac{d\rho}{d\vp}
\,+\,\frac{db}{d\si}\, \frac{d\si}{d\vp}\Big)\,{d\vp}\,.
\label{FPdet}
\eeq
As we use the gauge condition
$\,b = \na_\mu B^\mu$, the scalar sector in (\ref{FPdet}) vanishes
and the procedure is not affected by the symmetry breaking. In
what follows, we assume the use of this kind of gauge fixing
and do not consider ``exotic'' options which involve scalar fields.

The second-order term in the expansion of the action is given by
\beq
S^{(2)}
\,\,=\,\,
-\, \frac{1}{2}\, \int_x
\big( \,\rh \,\,\, \si \,\, B^\mu \,\big)
\begin{pmatrix}
H_{11} & H_{12} & H_{13} \\
H_{21} & H_{22} & H_{23} \\
H_{31} & H_{32} & H_{33}
\end{pmatrix}
    \begin{pmatrix}
    \rh \\ \si \\ B_\nu
    \end{pmatrix} ,
\eeq
where the matrix operator $\hat{H}$ is hermitian. The components are
\beq
\begin{aligned}
H_{11}
&\, = \,\Box + m^2 - \xi R - g^2 A^2 + \frac{\la}{2} \,\ph^2
+ \frac{\la_{12}}{6} \, \chi^2 \, ,
\\
H_{12} &\, = \,2g A^\mu \na_\mu + g\,(\na A)
+ \frac{\la_{12}}{3} \,\ph \chi \, ,
\\
H_{13} &\, = \,-2g^2A^\nu\ph + 2g\,(\na^\nu\chi) + g\chi\na^\nu \,,
\\
H_{21} &\, = \,-2g A^\mu \na_\mu - g\,(\na A)
+ \frac{\la_{12}}{3}\, \ph \chi \, ,
\\
H_{22} &\, = \,\Box + M^2 - \Xi R - g^2 A^2 + \frac{\la}{2} \,\chi^2
+ \frac{\la_{12}}{6} \, \ph^2 \,,
\\
H_{23} &=- 2g^2A^\nu\chi - 2g\,(\na^\nu\ph) - g\,\ph\na^\nu \,,
\\
H_{31} &=- 2g^2A_\mu\ph + g\,(\na_\mu \chi) - g\chi\na_\mu \,,
\\
H_{32} &=- 2g^2A_\mu\chi - g\,(\na_\mu \ph) + g\ph\na_\mu\, ,
\\
H_{33} &\, = \,-\de_\mu^\nu\Box + R_\mu^\nu
- \de_\mu^\nu g^2\,(\ph^2 + \chi^2)\,.
\end{aligned}
\eeq
The matrix $\hat{H}$ can be rewritten as
\beq
\hat{H}
\,=\,
    \begin{pmatrix}
+1 & 0 & 0 \\
0 & +1 & 0 \\
0 & 0 &-1
\end{pmatrix} \hat{H'}\,.
\eeq
The first factor $\diag(1,\,1,\,-1)$ does not contribute to the
divergences and $\hat{H}'$ has the form
\beq
\hat{H}'\, = \,\hat{1}\Box + 2 \hat{h}^\al\,\na_\al + \hat{\Pi} \, ,
\label{Hprime}
\eeq
such that one can employ the standard Schwinger-DeWitt
technique \cite{dewitt1965,dewitt2003} for computing the
divergences of $\Tr\log \hat{H}'$. The elements of the operator
(\ref{Hprime}) have the form
\beq
&&
2\hat{h^\al}
\, =\,
\begin{pmatrix}
0 & 2gA^\al & g\chi g^{\al\nu} \\
- 2gA^\al & 0 & - g\ph g^{\al\nu} \\
g\de^\al_\mu\chi & - g\de^\al_\mu\ph & 0
\end{pmatrix}
\quad
\mbox{and}
\,\,\quad
\hat{\Pi}
\,=\,
\left(
\begin{array}{ccc}
\Pi_{11}  &  \Pi_{12}  &  \Pi_{13} \\
\Pi_{21}  &  \Pi_{22}  &  \Pi_{23} \\
\Pi_{31}  &  \Pi_{32}  &  \Pi_{33}
\end{array}
\right),
\qquad
\label{hPi}
\eeq
where
\beq
&&
\Pi_{11} \,=\,
m^2 - \xi R - g^2 A^2 + \frac{\la}{2} \ph^2 + \frac{\la_{12}}{6} \chi^2,
\nn
\\
&&
\Pi_{12} \,=\,
\frac{\la_{12}}{3} \ph \chi + g(\na A),
\nn
\\
&&
\Pi_{13} \,=\,
-2g^2A^\nu\ph + 2g(\na^\nu\chi),
\nn
\\
&&
\Pi_{21} \,=\,
\frac{\la_{12}}{3} \ph \chi - g(\na A),
\nn
\\
&&
\Pi_{22} \,=\,
M^2 - \Xi R - g^2 A^2 + \frac{\la}{2} \chi^2 + \frac{\la_{12}}{6} \ph^2,
\nn
\\
&&
\Pi_{23} \,=\, -2g^2A^\nu\chi - 2g(\na^\nu\ph),
\nn
\\
&&
\Pi_{31} \,=\,
2g^2A_\mu\ph - g(\na_\mu\chi),
\nn
\\
&&
\Pi_{32} \,=\,
2g^2A_\mu\chi + g(\na_\mu\ph),
\nn
\\
&&
\Pi_{33} \,=\,
-R_\mu^\nu + \de^\nu_\mu g^2(\ph^2 + \chi^2).
\eeq
The one-loop contribution to the effective action is given by
\beq
    \bar{\Ga}^{(1)}_{div}
    \, =\,
    \frac{i}{2} \Tr \log \hat{H}'
    \, -\,
    i \Tr \log \hat{H}_{gh} ,
\eeq
where $\hat{H}_{gh}$ is the operator corresponding to the ghost action,
\beq
\hat{H}_{gh}\,=\, \Box \, .
\eeq
Using the Schwinger-DeWitt technique, the divergent part of the
one-loop effective action is given by \cite{dewitt1965}
\beq
&&
\bar{\Ga}^{(1)}_{div}
\,\,=\,\,
-\, \frac{\mu^{n-4}}{\vp}\,\int d^n x\,\sqrt{-g}
\,\,\tr \bigg\{\frac{\hat{1}}{360}\,\big(3C^2 - E_4 + 2\Box R\big)
 \nn
 \\
&&
\qquad
\qquad
+ \,\,
\frac{1}{2}\,\hat{P}^2 + \frac{1}{12}\,\hat{\mathcal{S}}_{\al\be}^2
+ \frac{1}{6}\,\Box \hat{P}
\bigg\},
\label{tracoa2}
\eeq
where $\mu$ is the renormalization scale, $\vp = 4\pi^2(n-4)$ is the
dimensional regularization parameter,
$C^2 = C^{\al\be\mu\nu}C_{\al\be\mu\nu}$ is the square of the Weyl
tensor, $E_4$ is the Gauss-Bonnet topological term,  The operators
$\hat{P}$ and $\hat{\mathcal{S}}_{\al\be}$ in (\ref{tracoa2}) are
given by
\beq
&&
\hat{P}
\,=\,
\hat{\Pi} + \frac{\hat{1}}{6}\,R - \na_\al \hat{h}^\al - \hat{h}_\al\hat{h}^\al,
\label{eqP}
\eeq
\beq
&&
\hat{\mathcal{S}}_{\al\be}
\,=\,
\hat{\mathcal{R}}_{\be\al}
+ \na_\be\hat{h}_\al - \na_\al\hat{h}_\be
+ \hat{h}_\be \hat{h}_\al - \hat{h}_\al \hat{h}_\be,
\label{eqS}
\eeq
where $\hat{\mathcal{R}}_{\al\be} = 0$ in the scalar sector and
$\hat{\mathcal{R}}_{\al\be} = [\mathcal{R}_{\al\be}]^\nu_{\,\,\mu}
= R^\nu_{\,.\,  \mu\al\be}$ for the vector field.

The contribution of the ghosts goes only to the vacuum (pure
metric-dependent) term $\,\bar{\Ga}^{(1)}_{div,\,vac}$, given by the
sum of the contribution of gauge vector and two real scalars. These
expressions are well-known (see, e.g., \cite{OUP}) and we can skip
them. In the matter sectors, the divergent part of the one-loop
effective action is given by the corresponding part of the general
expression (\ref{tracoa2}).  Some useful technical details can be
found in Appendix A. The total derivative terms are included for
completeness.
\beq
&&
\bar{\Ga}^{(1)}_{div}
\, \, = \, \,
-\, \frac{\mu^{n-4}}{\vp}
\int_{x,\,n} 
\bigg\{
 -2 g^2 (\na \ph)^2
 -2 g^2 (\na \chi)^2
 + 4 g^3 A^\mu\big[\ph (\na_\mu \chi) - \chi (\na_\mu \ph)\big]
\nn
\\
&&
- \,\,2 g^4 A^2 (\ph^2+\chi^2)
\,-\,\frac16\,g^2 F_{\mu\nu}^2
\,-\, \frac13\, g^2 R (\ph^2 + \chi^2)
- \,(m^2 - \tilde{\xi}R)g^2\chi^2
 \nn
 \\
 &&
- \,(M^2- \tilde{\Xi}R)g^2\ph^2
+ (m^2 - \tilde{\xi}R)\Big(\frac{\la}{2}\ph^2 + \frac{\la_{12}}{6}\chi^2\Big)
+ (M^2 - \tilde{\Xi}R)\Big(\frac{\la}{2}\chi^2 + \frac{\la_{12}}{6}\ph^2\Big)
\nn
\\
&&
+ \,\Big(\frac{\la^2}{8}+\frac{\la_{12}^2}{72} - \frac{\la_{12}}{6}  g^2
+ g^4\Big)\big(\ph^2 + \chi^2\big)^2+\Big[\frac{\la\la_{12}}{6}
- \frac{\la^2}{4}+\frac{\la_{12}^2}{12}+(\la_{12} - \la)g^2\Big]\ph^2\chi^2
\nn
\\
&&
 \qquad \qquad
+ \, \Big(\frac{g^2}{3}+\frac{\la}{12}
 + \frac{\la_{12}}{36} \Big)\Box(\ph^2 + \chi^2)
\bigg\}\,\,+\,\,\bar{\Ga}^{(1)}_{div,\,vac},
\label{effec}
\eeq
where $\tilde{\xi} = \xi - \frac{1}{6} $ and
$\tilde{\Xi} = \Xi - \frac{1}{6} $. The one-loop counterterms maintain
the structure of the classical action (\ref{acao2esc}) that guarantees the
gauge invariance.  The last means that, at the one-loop level, the theory
with the broken symmetry is renormalizable.

\section{Renormalization group equations}
\label{sec5}

It proves useful to consider all three types of symmetry breaking,
i.e., by $\la_{12} \neq \la$, by $\Xi \neq \xi$, and by $M \neq m$,
at the same time. Later on, they can be separated for the physical
analysis. Thus, we start with the full list of renormalization group
equations.

The renormalization relations for the fields are as follows
\beq
&&
\ph_0 \,=\, \mu^{\frac{n-4}{2}}\Big(1-\frac{2g^2}{\vp}\Big)\ph \, ,
\nn
\\
&&
\chi_0 \,=\,\mu^{\frac{n-4}{2}}\Big(1-\frac{2g^2}{\vp}\Big)\chi\, ,
\nn
\\
&&
A^0_\mu \,=\,\mu^{\frac{n-4}{2}}\Big(1+\frac{g^2}{3\vp}\Big) A_\mu \,.
\label{renfields}
\eeq
For the coupling constants, including using the relation
\ $g_0A^0_\mu= gA_\mu$ \ to guarantee the gauge invariance,
we get
\beq
&&
g_0 \,=\, \mu^{\frac{4-n}{2}}\Big(1-\frac{g^2}{3\vp}\Big)g\, ,
\nn
\\
&&
\la_0 \,=\, \mu^{4-n}\Big(\la - \frac{3\la^2}{\vp}
- \frac{\la_{12}^2}{3\vp} + \frac{4\la_{12}}{\vp} g^2
- \frac{24g^4}{\vp}+\frac{8\la g^2}{\vp}\Big)\,,
\nn
\\
&&
\la^0_{12} \,=\,\mu^{4-n}\Big(\la_{12}
- \frac{2\la\la_{12}}{\vp}-\frac{4\la_{12}^2}{3\vp}
+ \frac{12\la g^2}{\vp} - \frac{8\la_{12}g^2}{\vp}
- \frac{24g^4}{\vp}+\frac{8\la_{12}g^2}{\vp}\Big)\, .
\label{rencoulings}
\eeq

Finally, for the masses and nonminimal parameters, we have
\beq
&&
m_0^2 \,=\, \Big(m^2 - \frac{\la}{\vp}m^2
+ \frac{4g^2}{\vp}m^2 -\frac{\la_{12}}{3\vp}M^2
+ \frac{2g^2}{\vp}M^2 \Big)\, ,
\nn
\\
&&
M_0^2 \,=\, \Big(M^2 - \frac{\la}{\vp}M^2+\frac{4g^2}{\vp}M^2
-\frac{\la_{12}}{3\vp}m^2 + \frac{2g^2}{\vp}m^2 \Big)\,,
\nn
\\
&&
\xi_0 \,=\,\xi+\Big[ \Big(\frac{4g^2}{\vp}
- \frac{\la}{\vp}\Big)\Big(\xi-\frac{1}{6}\Big)
- \Big(\frac{\la_{12}}{3\vp} - \frac{2g^2}{\vp}\Big)
\Big(\Xi-\frac{1}{6}\Big)  \Big]\,,
\nn
\\
&&
\Xi_0 \,=\,\Xi+\Big[ \Big(\frac{4g^2}{\vp}
- \frac{\la}{\vp}\Big)\Big(\Xi-\frac{1}{6}\Big)
- \Big(\frac{\la_{12}}{3\vp} - \frac{2g^2}{\vp}\Big)
\Big(\xi-\frac{1}{6}\Big)  \Big]\,.
\label{renxis}
\eeq
The last relations satisfy the usual hierarchy, in the sense
(\ref{renxis}) depend on the (\ref{rencoulings}), but not the
opposite.
Now we can derive the beta and gamma functions as
\beq
\be_P\,=\,\lim_{n\to \,4} \mu\frac{dP}{d\mu}
\qquad
\mbox{and}
\qquad
\ga_H\,=\,\lim_{n\to \,4} \mu\frac{dH}{d\mu}\,,
\label{betagama}
\eeq
where $P = \big(g, \la, \la_{12}, m^2, M^2, \xi, \Xi\big)$
are the renormalized parameters and $H = \big(\phi, \chi, A_\mu\big)$
are renormalized fields. Using the renormalization relations, we obtain
the beta functions
\beq
&&
\be_g \,=\, \frac{g^3}{3(4\pi)^2} \, ,
\nn 
\\
\nn 
&&
\be_{\la_{12}} \,=\, \frac{1}{(4\pi)^2}
\Big(\frac{4}{3}\la_{12}^2+2\la\la_{12}-12\la g^2 + 24g^4\Big) \, ,
\\
\nn 
&&
\be_\la
\,=\,\frac{1}{(4\pi)^2}\Big(3\la^2  +\frac{\la_{12}^2}{3}
- 8\la g^2-4\la_{12}g^2 + 24g^4\Big) \, ,
\\
&& \be_{m^2}
\,=\, \frac{1}{(4\pi)^2}\Big(\la m^2 + \frac{\la M^2}{3}-2g^2M^2
- 4g^2m^2\Big) \, ,
\nn
\\
&&
\be_{M^2}
\,=\, \frac{1}{(4\pi)^2}\Big(\la M^2 + \frac{\la m^2}{3}
-2g^2m^2-4g^2M^2\Big) \,,
\nn 
\\
&&
\be_\xi \,=\, \frac{1}{(4\pi)^2}\Big[
\Big(\xi-\frac{1}{6}\Big)\big(\la-4g^2\big) + \Big(\Xi-\frac16\Big)
\Big(\frac{\la_{12}}{3}-2g^2\Big)\Big] \, ,
\nn 
\\
&&
\be_\Xi \,=\, \frac{1}{(4\pi)^2}\Big[
\Big(\Xi-\frac16\Big)\big( \la - 4g^2 \big)
+ \Big(\xi-\frac16\Big)\Big(\frac{\la_{12}}{3}-2g^2\Big)\Big] \, .
\label{betas}
\eeq
The gamma functions are
\beq
\ga_{\ph} \,=\,\ga_{\chi}\,=\,\frac{g^2}{(4\pi)^2}
\qquad
\mbox{and}
\qquad
\ga_A \,=\, -\,\frac{g^2}{3(4\pi)^2}\,.
\label{gammas}
\eeq
The derivation of the formulas (\ref{betas}) and (\ref{gammas})
is pretty standard, i.e., based on
the renormalization relations (\ref{renfields}),
(\ref{rencoulings}) and (\ref{renxis}). In the dimensional
regularization, the factor $1/(4\pi)^2$ comes from the
definition of $\vp$ in (\ref{tracoa2}).

Let us note that the expressions (\ref{betas}) satisfy the usual
tests related to local conformal invariance in the massless limit.
On top of this, after taking the limit (\ref{back}) we arrive at the
usual renormalization group functions of sQED in both (\ref{betas})
and (\ref{gammas}) cases.

The gamma function for the $A_\mu$ and the beta function for
$g$ does not change under the limit (\ref{back}). In particular,
for the coupling constant $g$ we get the conventional running
\beq
g^2 (t) \,=\, g_0^2 \,\bigg[1 - \frac{2g_0^2\, t}{3(4\pi)^2}\bigg]^{-1},
\label{gt}
\eeq
where $t \,=\, \log\big(\mu / \mu_0\big)$ and $g_0 = g (\mu_0)$.
As usual for the Abelian models, we can consistently explore
the running only in the IR (low-energy limit). On the other hand,
in the IR the presence of masses implies the decoupling and the
running stops below the corresponding energy threshold. Thus,
let us consider only the massless limit and pay special attention
to the running of the scalar couplings. In the massless theory only
these two effective charges and the nonminimal parameters
define the gauge symmetry breaking in the theory (\ref{acaoclas2}).

To explore the behavior of the scalar couplings, we follow the
approach elaborated in the non-abelian gauge theories (see,
e.g., \cite{ChengEichtenLi1974}). Introduce the new variables
\beq
\bar\lambda(t)
\,=\, \frac{\la(t)}{g^2(t)}
\qquad
\mbox{and}
\qquad
\bar\la_{12}(t) = \frac{\la_{12}(t)}{g^2(t)}\,,
\label{barred}
\eeq
that provides the equations
\beq
\label{barlam12}
\frac{d\bar{\la}_{12}}{d\tau}
\,\,=\,\, \frac43\,\bar{\la}_{12}^2
\,+\, 2 \bar{\la}_{12} \Big(\bar{\la}
\, -\, \frac{1}{3}\Big)
\,-\, 12\bar{\la} + 24
\eeq
and
\beq
\label{barlam}
\frac{d\bar\lambda}{d\tau}
\,\,=\,\,
3\Big(\bar{\la} - \frac{13}{9}\Big)^2
\,+\, \frac13\,\Big(\bar{\la}_{12} - 6\Big)^2
\,+\, \frac{155}{27}\,,
\eeq
where \
\beq
\tau \,=\, -\,\frac{3}{2}\,\ln\Big[1-\frac{2g_0^2\, t}{3(4\pi)^2}\Big].
\label{tau}
\eeq
In the IR, when $\,t \to -\infty$, the new variable behaves as
$\,\tau \to -\infty$.

At this point, we can make a simple test. Let us remember that
by setting $\la_{12} = \la$ or, equivalently,
$\bar{\la}_{12} = \bar{\la}$, we recover gauge symmetry in the
interaction term. In this case, the splitting of the couplings should
disappear. Taking this limit according to (\ref{back}), the two
equations (\ref{barlam12}) and (\ref{barlam}) coincide. This is
a good (albeit partial) verification of the correctness of the
calculations.
Introducing the new parameter for the difference
\beq
\label{la_la12}
\ep(\tau) \,=\, \bar{\la}_{12}(\tau) \,-\, \bar{\la}(\tau)\,,
\eeq
we arrive at the differential equation
\beq
\frac{d\epsilon}{d\tau}
\,=\,
\ep \Big(\ep + 4\bar{\la} + \frac{10}{3}\Big).
\label{beep}
\eeq
We can explore the equations assuming that the deviation is
small, i.e., $|\ep| \ll \bar{\la}$.
In this regime, $\bar{\la} \approx \bar{\la}_{12}$ and the
\textit{r.h.s.}'s of both (\ref{barlam12}) and (\ref{barlam})
become
\beq
\frac{10}{3}\,\bar{\la}^2 \,-\,\frac{38}{3}\,\bar{\la} \,+\, 24.
\label{betalaco}
\eeq
This expression does not have real roots, which is clear already
from Eq.~(\ref{barlam}). This means that in the Abelian model
with scalar fields this approach does not work because there is
no consistent IR (neither UV) asymptotic behavior for
$\bar{\la}(\tau)$.

In order to have some conclusive output, consider an approximation
with the hierarchy of coupling constants $g^2 \ll \la$ and omit
all terms with $g^2$. In this case, the $\la(t)$ approaches zero in the
IR (which means, $\mu \to 0$ or $t \to -\infty$) as
\beq
\la (t) \,=\, \la_0 \,\big[1 - b^2\la_0\, t\big]^{-1},
\qquad
b^2 = \frac{10}{3(4\pi)^2}\,.
\label{lat}
\eeq
The equation for the difference $\ep(t) = \la_{12}(t) - \la(t)$ has
the form
\beq
\frac{d\ep}{dt}
\,=\,
 \frac{1}{(4\pi)^2}\,\ep \,(\la_{12} + 3 \la).
\label{ept}
\eeq
Assuming $\ep$ small compared to both scalar couplings,  we consider
$\la_{12} = \la$ and easily arrive at the approximate solution
\beq
\ep(t) \,=\, \ep_0 \big[1 - \la_0 b^2\, t\big]^{- \ka},
\label{epsol}
\eeq
where $\ka = 6/5$.
The solution (\ref{epsol}) indicates that if the gauge symmetry is
broken by the weak violation $\la_{12} \neq \la$, at the scale
$\mu_0$, the symmetry tends to be restored in the IR, when
$\mu \ll \mu_0$ and, asymptotically, $t \to -\infty$. Regardless
the running is logarithmic, the solution indicates, in the given
approximations, restoration of the gauge symmetry asymptotically
in the IR. It would be interesting to apply the explicit symmetry
breaking scheme to non-abelian theories, where one can explore
a more interesting UV limit.

\section{Conclusions and discussions}
\label{sec6}

We have presented a model with Abelian gauge symmetry, which
is equivalent to the usual complex (charged) scalar field theory,
but is formulated in terms of two real scalar fields in curved
spacetime. The standard gauge transformations of the complex
scalar can be reformulated for two real scalar fields, to ensure
the gauge invariance.
It is interesting that one cannot extent the
Abelian scalar model by introducing more real scalars. This
means, the generalizations of the model may be achieved only
using the non-abelian symmetry.

Introducing two different masses for the two real scalar fields leads
to the qualitatively new way of \textit{explicit} symmetry breaking
in the classical action, that does not require the Higgs mechanism.
A similar effect can be achieved by modifying the self-interaction
term for the two scalars, or even by splitting the parameters of
nonminimal interaction of the scalars with curvature. The derivation
of the one-loop counterterms has shown that the new theory with
broken symmetry is renormalizable at least at the one-loop level.
This calculation was done by using the standard Schwinger-DeWitt
technique.

The renormalizability of the theory with explicitly broken Abelian
gauge symmetry enables one using the renormalization group
method to explore the running of the symmetry-violating
parameters. However, since the UV limit in the Abelian model meets
known problems with the Landau pole, one can consistently explore
such a running only in the IR limit. This rules out the use of
massive theories because in the IR we expect the low-energy
decoupling and the running should stop. Thus, we consider only the
IR running for self-coupling in a massless theory. It turns out
that the splitting of the self-couplings tends to disappear in the IR
limit, which means a restoration of the gauge symmetry.

At the moment, we do not know any physical application of the
new way of gauge symmetry breaking.
From the mathematical side, we note that it looks possible to
generalize the new symmetry breaking scheme to the non-abelian
theory, e.g., for the case of complex scalars in the fundamental
representation of the $SU(2)$ or a more general symmetry group.
In this case, one can expect the renormalizability of the theory
with broken symmetry, at least at the one-loop level. This would
open the way for exploring the UV running of the
symmetry-breaking parameters, including in the massive models.

It is unclear whether the explicit breaking of symmetry my splitting
masses of real scalars, or splitting the self-scalar coupling sector, or
splitting the nonminimal parameters of the interaction with scalar
curvature, may be useful. At the present state of art, the new way of
symmetry breaking is a kind of a theoretical curiosity. However,
it is possible to imagine extensions with potentially interesting
applications, such as to fermions or to the non-abelian symmetry.
In view of this possibility, it makes sense to see to which extent the
new simple model described above may be theoretically consistent
beyond one loop order. At the moment, we do not have conclusive
answer to this question, but some preliminary considerations are
collected in Appendix B.

\section*{Acknowledgements}
The authors are grateful to Riccardo Feccio for a useful correction
concerning the analysis of the renormalization group equations.
I.Sh. is grateful to CNPq (Conselho Nacional de Desenvolvimento
Cient\'{i}fico e Tecnol\'{o}gico, Brazil)  for the partial support
under grant 305122/2023-1.

\section*{Appendix A. Intermediate Formulas for One-Loop Divergences}

In this appendix, we present the intermediate formulas for the
calculation of one-loop divergences. Using \eqref{eqP} and
\eqref{eqS}, the operators $\hat{P}$ and $\hat{S}_{\al\be}$
are given by
\beq
    &&
        \hat{P}
        \,= \,
        P^\nu_\mu
        \, = \,
        \begin{pmatrix}
       P_{11}& P_{12} & P_{13} \\
       P_{21} & P_{22} & P_{23}\\
       P_{31} & P_{32} & P_{33}
        \label{operatorP}
    \end{pmatrix},
\eeq
with the elements
\beq
&&
P_{11} \,=\, m^2 - \tilde{\xi} R  + \frac{\la}{2} \ph^2
+ \frac{\la_{12}}{6} \chi^2 - \frac{1}{4}g^2\chi^2\de^\nu_\mu \,,
\nn
\\
&&
P_{12} \,=\,P_{21} \,=\, \frac{\la_{12}}{3} \ph \chi
+ \frac{1}{4}g^2\chi\ph\de^\nu_\mu\,,
\nn \\
&&
P_{13} \,=\, \frac32\,g \big[(\na^\nu\chi) - gA^\nu\ph\big]\,,
\qquad
P_{21} \,=\, \frac{\la_{12}}{3} \ph \chi
+ \frac{1}{4}g^2\chi\ph\de^\nu_\mu\,,
\nn
\\
&&
P_{22} \,=\, M^2 - \tilde{\Xi} R  + \frac{\la}{2} \chi^2
+ \frac{\la_{12}}{6} \ph^2 - \frac{1}{4}g^2\ph^2\de^\nu_\mu \, ,
\nn
\\
&&
P_{23} \,=\, -\frac{3}{2}g \big[(\na^\nu\ph) + gA^\nu\chi\big] \, ,
\qquad
P_{31} \,=\, -\frac{3}{2}g \big[(\na^\nu\chi) - gA^\nu\ph \big]\, , \nn \\
&&
P_{32} \,=\, \frac{3}{2}g\big[(\na^\nu\ph) + gA^\nu\chi \big]\, ,
\,\,\,\,\qquad
P_{33} \,=\, \frac{1}{6}R\de^\nu_\mu
- R_\mu^\nu + \frac{3}{4}\de^\nu_\mu g^2\big(\ph^2 + \chi^2\big)\,,
\eeq
where $\tilde{\xi} = \xi - \frac{1}{6} $ and $\tilde{\Xi} = \Xi - \frac{1}{6} $.
The operator $\hat{\mathcal{S}}_{\al\be}$ is defined as
\beq
    &\hat{\mathcal{S}}_{\al\be} = [S_{\al\be}]^\nu_\mu = \begin{pmatrix}
    s_{11} & s_{12} & s_{13} \\
    s_{21} & s_{22} & s_{23} \\
    s_{31} & s_{32} & s_{33}
    \end{pmatrix},
\eeq
where the components are given by
\beq
&&
s_{11} \,=\,  \frac{1}{4}\,g^2\chi^2 \big(g_{\al\mu}\de^\nu_\be
- g_{\be\mu}\de^\nu_\al\big)\,,
\nn
\\
&&
s_{12} \,=\, -\frac{1}{4}\,g^2\ph \chi \big(g_{\al\mu}\de^\nu_\be
- g_{\be\mu}\de_\al^\nu\big) - gF_{\al\be}\,,
\nn
\\
&&
s_{13} \,=\, -\frac{1}{2}\,g^2\ph \big(A_\be \de^\nu_\al
- A_\al \de^\nu_\be\big)
+ \frac{1}{2}\,g \big[(\na_\be\chi) \de^\nu_\al
- (\na_\al\chi) \de^\nu_\be\big] \,,
\nn
\\
&&
s_{21} \,=\, -\frac{1}{4}\,g^2\ph \chi \big(g_{\al\mu}\de^\nu_\be
- g_{\be\mu}\de_\al^\nu\big) + gF_{\al\be}\,,
\nn
\\
&&
s_{22} \,=\, \frac{1}{4}\,g^2\ph^2 \big(g_{\al\mu}\de^\nu_\be
- g_{\be\mu}\de^\nu_\al \big)\,,
\nn
\\
&&
s_{23} \,=\, -\frac{1}{2}\,g^2\chi \big(A_\be \de^\nu_\al
- A_\al \de^\nu_\be\big)
-\frac{1}{2}\,g \big[(\na_\be\ph) \de^\nu_\al
- (\na_\al\ph) \de^\nu_\be \big]\,,
\nn
\\
&&
s_{31} \,=\, \frac{1}{2}\,g^2\ph
\big( A_\al g_{\be\mu} - A_\be g_{\al\mu}\big)
+ \frac{1}{2}\,g
\big[(\na_\be\chi)g_{\al\mu} - (\na_\al\chi)g_{\be\mu}\big]\,,
\nn
\\
&&
s_{32} \,=\, \frac{1}{2}\,g^2\chi \big( A_\al g_{\be\mu}
- A_\be g_{\al\mu}\big)
- \frac{1}{2}\,g \big[(\na_\be\ph)g_{\al\mu}
- (\na_\al\ph)g_{\be\mu}\big]\,,
\nn
\\
&&
s_{33} \,=\, \frac{1}{4}\,g^2 \big(\chi^2
+ \ph^2) (g_{\be\mu} \de^\nu_\al
- g_{\al\mu} \de^\nu_\be\big) - R^\nu_{\,.\,  \mu\al\be}\,.
\label{commutador}
\eeq

Replacing these operators in the general formula (\ref{tracoa2}),
we arrive at the expression for one-loop divergences (\ref{effec}),

\section*{Appendix B.
On the renormalizability in the higher loop order}

We have found that the theory with explicitly broken symmetry
remains renormalizable at the one-loop level. One can be curious
whether there is a chance to meet renormalizability in the next
orders. The proof of this feature is a nontrivial task because
there is no symmetry protecting the quantum theory from the
qualitatively new divergences.
Let us start saying that we do not have a definite answer to
this question and can only present some preliminary considerations
for the simplest version of symmetry breaking.

Since the theory remains power-counting renormalizable,
the list of the possible divergences which might violate the
multiplicative renormalizability are all in the vector sector. The
candidate terms include the following:

\textit{(i)} Longitudinal divergence $(\pa_\mu A^\mu)^2$. By
dimensional reasons, this
term may appear only owing to the splitting of the self-scalar
coupling $\la$ into a couple $(\la,\,\la_{12})$.

\textit{(ii)} The divergence with the vector field mass
$M_v^2 A_\mu A^\mu= M_v^2A^2$. This term may appear
because of the splitting of $\la$ and/or because
of the splitting of the scalar masses.

\textit{(iii)} Nonminimal divergences $A^\mu A^\nu R_{\mu\nu}$ and
$R A_\mu A^\mu$, which were recently discussed in \cite{DerVecScal}.
These terms may appear because of the splitting of the self-scalar
coupling $\la$, or because of the splitting of $\xi$ into the pair of
nonminimal parameters $(\xi,\,\Xi)$.

It is clear that the simplest case is \textit{(ii)} because it might
come only from the single source. Let us discuss only this particular
case, assuming the background metric is flat and the symmetry
breaking is only because of the splitting of masses.
In case of $m_1 = m_2$, there are
no $M^2A^2$-divergences because of the gauge symmetry.
Thus, the \textit{(ii)}-type $n$-loop divergences are proportional
to $m_1-m_2$. Taking the power counting into account, the
coefficient should be $(m_1 - m_2)(a m_1 + b m_2)$, where
$a$ and $b$ are some dimensionless numerical coefficients.
On the other hand, since there is a duality under
$m_1 \leftrightarrow m_2$, we get $b=-a$ and the combination
can be only $(m_1 - m_2)^2$.

Consider $n$-loop approximation, assuming there are no
\textit{(ii)}-type divergences at the  $(n-1)$-loop approximation.
Then the $M_v^2A^2$-type divergences can only arise from the
last integration. Since we already know that the one-loop
approximation is renormalizable, we can consider two-loop
diagrams and get an idea of what takes place for any $n$.
The classical vertices of the theory with explicitly broken symmetry
(by masses only) are shown in Fig.~\ref{Fig1} and the one-loop
diagrams of our interest in Fig.~\ref{Fig2}. Until this point, we
know that the \textit{(ii)}-type divergences are not generated. At
the two-loop level we meet three types of diagrams from
Figs.~\ref{Fig3}.

\begin{figure}[ht!]
\centering
\includegraphics[width=0.4\textwidth]{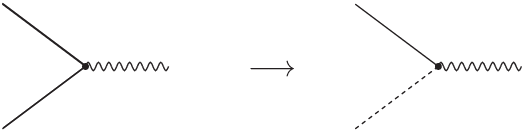}
\vskip 3mm

$\,$\qquad
\qquad
\includegraphics[width=0.53\textwidth]{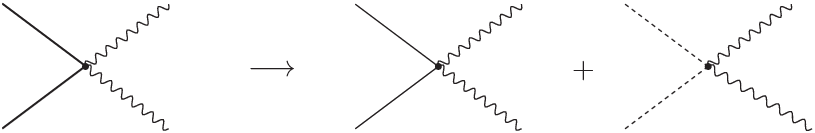}
\begin{quotation}
\vspace{-5mm}
\caption{\small
Transformation of the vertices under splitting one complex scalar
into two real scalars with different masses. Here and in what follows,
the bold line indicates complex scalar. }
\label{Fig1}
\end{quotation}
\end{figure}

\begin{figure}[ht!]
\centering
\includegraphics[width=0.77\textwidth]{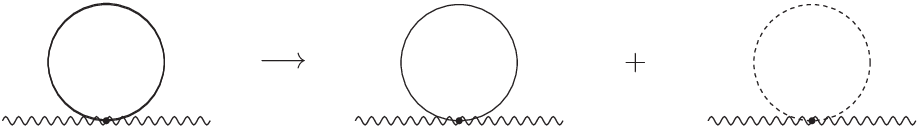}
\vskip 4mm

\includegraphics[width=0.54\textwidth]{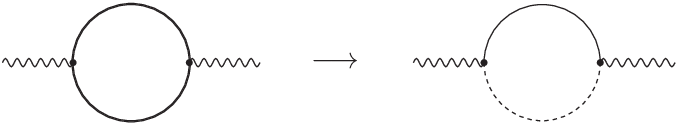}
\begin{quotation}
\vspace{-5mm}
\caption{\small
Transformation of the snail and bubble one-loop diagrams under
splitting a complex scalar into two real scalars with different
masses.}
\label{Fig2}
\end{quotation}
\end{figure}
\vskip 3mm

\begin{figure}[ht!]
\centering

\includegraphics[width=0.7\textwidth]{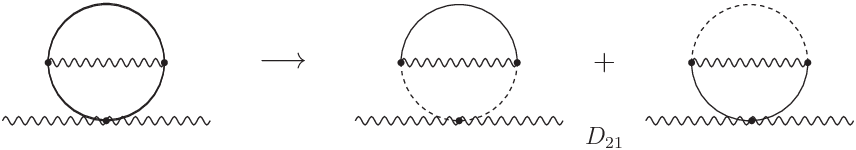}
\vskip 4mm

\includegraphics[width=0.8\textwidth]{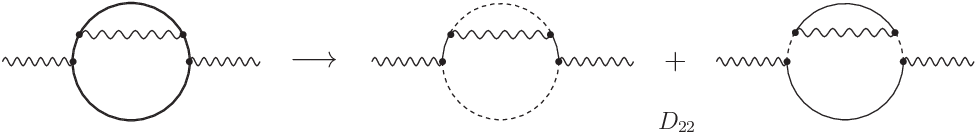}
\vskip 2mm

\includegraphics[width=0.52\textwidth]{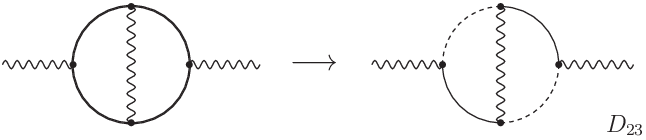}
\begin{quotation}
\vspace{-5mm}
\caption{\small
Transformation of the two-loop diagrams under splitting a complex
scalar into two real scalars with different masses.
\ First line $D_{21}$ represents the snail-type diagrams;
\ second line $D_{22}$ represents the first type two-loop bubble
diagrams
\ third line $D_{23}$ represents the second type two-loop bubble
diagrams.}
\label{Fig3}
\end{quotation}
\end{figure}
\vskip 3mm

We note that each group of two-loop diagrams, i.e., $D_{21}$,
$D_{22}$ and $D_{23}$, individually may not be proportional
to $(m_1 - m_2)^2$ since $M^2A^2$ may not cancel for each
group in the case of unbroken symmetry. However, cancellation
is guaranteed for the sum. This feature holds in all loop orders
for the superficial integration. The fact each group of diagrams
individually is proportional to $(m_1 - m_2)^2$ reflects the
relations between the diagrams of the theory with unbroken
symmetry.
If each group (e.g., $D_{21}$, $D_{22}$ and $D_{23}$ in the
two-loop case) is free from
$M^2A^2$-divergences in the symmetric phase, then the factor
$(m_1 - m_2)^2$ should not be present in each of the groups of
the diagrams separately, in order to achieve an overall $M^2A^2$
cancellation in the symmetric version. Certainly, these
considerations cannot guarantee the absence of  $M^2A^2$
divergences at higher loops and the question remains open.


\end{document}